\documentclass[seceq]{ptptex}
\usepackage{wrapft}

\usepackage{graphicx}

\def\e{\kern+.6ex\lower.42ex\hbox{$\scriptstyle \iota$}\kern-1.20ex e}

\begin{document}

\markboth{
A. M.  Phyu {\em et al.}}{
The Complex Energy Method Applied to the Nd Scattering}

\title{
The Complex Energy Method Applied to the Nd Scattering \\ with a Model Three-Body Force}

\author{
Aye Mya \textsc{Phyu}$^{1,}$\footnote{E-mail: ayemyaphyu.phyu5@gmail.com
},
Hiroyuki \textsc{Kamada}$^{2,}$\footnote{E-mail: kamada@mns.kyutech.ac.jp
},
Jacek \textsc{Golak}$^{3,}$\footnote{E-mail: ufgolak@cyf-kr.edu.pl},
 Htun Htun \textsc{Oo}$^{4,}$\footnote{E-mail: htunhtun.oo93@googlemail.com},
Henryk~\textsc{Wita\l a}$^{3,}$\footnote{E-mail: ufwitala@cyf-kr.edu.pl},
Walter~\textsc{Gl\"ockle}$^{5,}$\footnote{E-mail: Walter.Gloeckle@tp2.ruhr-uni-bochum.de}
}

\inst{
$^1$ Department of Physics, Mandalay University, Mandalay, Union of Myanmar \\
$^2$Department of Physics, Faculty of Engineering, Kyushu Institute of Technology,
Kitakyushu 804-8550, Japan \\
$^3$M. Smoluchowski Institute of Physics, Jagiellonian University, PL-30059 Krak\'ow, Poland \\
$^4$Department of Physics, Meiktila University, Meiktila, Union of Myanmar \\
$^5$Institut f\"ur Theoretische Physik II, Ruhr-Universit\"at Bochum, D-44780 Bochum, Germany
}




\abst{
Using the complex energy method, the problem of nucleon-deuteron scattering is solved with a simple
three-body force having a separable form. 
Our results are compared with the results of modern direct two-variable calculations and a good agreement is found.
This forms a firm base for other applications of the complex energy method.
}

\maketitle

\section{Introduction}

Faddeev calculations of nucleon-deuteron (Nd) scattering with a
three-nucleon 
force (3NF) have been 
performed for many years\cite{Gloeckle0}. 
Modern nucleon-nucleon potential stem from the chiral effective 
field theory\cite{Epelbaum}.
This theory gives us also various 3NFs but for many years 
numerous calculations were performed
with the Tucson-Melbourne 3NF\cite{TM3NF}.
There are two  alternative formulations leading to the scattering amplitude
with 
3NF in the momentum 
space\cite{Hueber}.
One of them was introduced by Gl\"ockle and Brandenburg \cite{Gloeckle1}.
In this scheme the Alt-Grassberger-Sandhas (AGS) equation for the
elastic  scattering 
transition operator $U$ is written as
\cite{Gloeckle1}
\begin{eqnarray}
U=PG_0^{-1}+(1+P)t_4 +Pt_1 G_0U+(1+P)t_4G_0t_1G_0U , 
\label{Eq1}
\end{eqnarray}
where $P$, $t_1$, $G_0$ and $t_4$ are 
the permutation operator ($P\equiv P_{12}P_{23}+P_{13}P_{23}$),
the two-body t-matrix, the free Green's function and the t-matrix 
 generated by a 3NF, 
respectively.

The other approach has been used by the Bochum-Krak\'ow
group\cite{Gloeckle0} (BK).
Here the equation for $U$ has the following form:
\begin{eqnarray}
U=PG_0^{-1}+(1+P)V_4^{(1)}(1+P) +Pt_1 G_0U+(1+P)V_4^{(1)}(1+P)G_0t_1G_0U ,~~~ 
\label{BKEq1}
\end{eqnarray}
where $V_4^{(1)}$ is a part of 3NF which is symmetric under the exchange of nucleons 2 and 3 so the full 3NF, $V_4$,  is given as $V_4=V_4^{(1)}(1+P)$.
In practice in the BK approach, first the auxiliary equation for the three-body 
operator $T$ is solved, 
\begin{eqnarray}
T|\phi \rangle=t_1P|\phi \rangle +(1+t_1G_0)V_4|\phi \rangle +
t_1G_0PT|\phi \rangle + (1+t_1G_0)V_4G_0T|\phi \rangle \, ,
\label{BKeq}
\end{eqnarray}
 with the incoming momentum space state $|\phi \rangle$ composed of
the relative nucleon-deuteron motion and the deuteron wave function.
Using this $T$, the elastic transition operator $U$ is given as
\begin{eqnarray}
U|\phi\rangle =PG_0^{-1}|\phi\rangle+
PT|\phi \rangle +V_4|\phi\rangle + V_4G_0T|\phi \rangle .
\label{Eq4}
\end{eqnarray}
The consequence of employing Eq.(\ref{BKeq}) is that the t-matrix of
the 3NF, $t_4$,  does not appear in this formulation.
Contrary to that, the approach based on Eq.~(\ref{Eq1}) requires a calculation 
of $t_4$ for a given 3NF.
The formalism of Eq.~(\ref{Eq1}) has already been used with a simple separable 3NF\cite{Oryu}.
In that calculation, however, the t-matrix of the 3NF was approximated by taking $t_4 \approx V_4$. 
   It is the aim of the present investigation to show that both schemes,
   (\ref{Eq1}) and (\ref{BKeq})--(\ref{Eq4}), yield the same numerical 
results under a consistent separable approximation  for the underlying dynamics
and avoiding an unnecessary approximation $t_4 \approx V_4$ in the approach 
based on Eq.~(\ref{Eq1}). Additionally, when using that approach, 
we apply the complex energy method described in detail in \cite{Kamada}.
In that reference it was shown that the complex energy method can be successfully
employed for such realistic interactions like the CD~Bonn and Argonne V14 
potentials. 

    In Section 2 we show details of our formalism which leads to an 
analytical t-matrix of  3NF 
and the modified Amado-Lovelace (AL) equation.
In Section 3, using the Yamaguchi separable potential and a separable
3NF, 
we will demonstrate results for
the Nd elastic scattering cross section and the  neutron-deuteron 
scattering length. These results will be compared with the results of standard calculations using the general two-variable formulation \cite{Gloeckle0}.
In Section 4 we give a short summary.

\section{Modified Amado-Lovelace equation}
We first rewrite 
 Eq.(\ref{Eq1})  as \cite{Oryu}  
\begin{eqnarray}
\tilde U=t_1G_0 P t_1 + t_1 G_0(1+P)t_4G_0t_1 +t_1G_0P\tilde U+ t_1G_0 (1+P)t_4 G_0 \tilde U ,
\label{Utilde}
\end{eqnarray}
where $\tilde U$ is related to $U$ via
$\tilde U =t_1 G_0 U G_0 t_1$ 
and the t-matrix of the 3NF $t_4$ is the solution of the Lippmann-Schwinger equation:
$t_4=V_4+V_4G_0t_4 $.
Next  we introduce the so-called  separable approximation for the two-body and three-body   interactions.
We choose first the Yamaguchi formfactor, \cite{Yamaguchi} $g_c(p)$,  for the pure S-wave two-body potential:
\begin{eqnarray}
v_c(p,p')=-\lambda_c g_c(p)g_c(p') \equiv -\lambda_c {1 \over p^2+\beta^2} {1\over p'^2+\beta^2} ,
\end{eqnarray}
where $p'$ ($p$) is the magnitude of the initial (final) 
relative momentum within the two-nucleon subsystem 
and $\lambda_c$  is the coupling coefficient.
We choose the same value for the $\beta$ parameter (1.4488 fm$^{-1}$) both for the $^1$S$_0$ ($c=\phi$) and $^3$S$_1$ ($c=d$) states.
The coupling coefficient $\lambda_d$ for the $^3$S$_1$ state is taken as $\lambda_d=8\pi\beta(\alpha+\beta)^2/m$ 
 to get the experimental deuteron binding energy with $\alpha=$0.2316 fm$^{-1}$ ,
 and the other coupling coefficient $\lambda_\phi=8\pi\beta(\alpha_0-\beta)^2/m $  
for the $^1$S$_0$ state 
allows us to reproduce the nucleon-nucleon phase shift parameters with $\alpha_0=$-0.040 fm$^{-1}$. 
The Lippmann-Schwinger equation,
$ t_1=v+vG_0t_1$ ,
is explicitly given as:

\begin{eqnarray}
t_1(p,p';E)=v(p,p')+\int _0^\infty {v(p,p'') t_1(p'',p';E) \over E -p''^2/m +i\epsilon}{p''^2 dp''
\over 2\pi^2}.
\end{eqnarray}
where $m$ is the
nucleon mass ($1/m=$41.47 MeV fm$^{2}$).
It is well known\cite{Yamaguchi} that the t-matrix $t_1$ of a separable potential can be calculated analytically and has also 
a separable form ($t_1(p,p';E)=g(p)\tau(E)g(p')$).

In the case of the 3NF we assume the following S-wave phenomenological separable potential
\begin{eqnarray}
V_4(p,q,p'q')=\lambda_4 h(p,q)h(p',q')=\lambda_4 {1 \over (p^2+{3\over 4}q^2+\Lambda^2)^2} 
{1\over (p'^2+{3\over 4}q'^2+\Lambda^2)^2} ,
\label{V4}
\end{eqnarray}
where $q$ is the Jacobi momentum for the spectator particle
and $\lambda_4$ is the 3NF coupling
 coefficient.  The choice of the $\Lambda$ parameter will be discussed in Section 3.
Like the two-body t-matrix, also the 3NF t-matrix $t_4$ can be obtained analytically,
again in a separable form: $t_4(p,q,p',q';E)=h(p,q)\tau_4(E)h(p',q')$.
The derivation is presented in the Appendix.

Using these separable formulas for $t_1$ and $t_4$ leads to a modified AL equation:
\begin{eqnarray}
&&X_{c,c'}(q,q_0;E)= Z_{c,c'}(q,q_0;E) \cr 
&&+\sum _{c''}\int _0 ^\infty Z_{c,c''}(q,q'';E)\tau_{c''}(E-{3q''^2\over 4m})X_{c'',c'}(q'',q_0;E) 
{q''^2dq''\over 2\pi^2}.
\label{AL}
\end{eqnarray}
The amplitudes $X_{c,c'}$ and $ Z_{c,c'} $ are written as
\begin{eqnarray}
\langle  p,q, c| \tilde U | p',q',c' \rangle = 
g_c(p)\tau_c(E-{3q^2\over 4m})X_{c,c'}(q,q';E)\tau_{c'}(E-{3q'^2\over 4m})g_{c'}(p'),
\end{eqnarray}
\begin{eqnarray}
&&\langle  p,q, c| t_1G_0 P t_1 + t_1 G_0(1+P)t_4G_0t_1 | p',q',c'\rangle \cr
&&=g_c(p)\tau_c(E-{3q^2\over 4m})Z_{c,c'}(q,q';E)\tau_{c'}(E-{3q'^2\over 4m})g_{c'}(p')\cr
&&=g_c(p)\tau_c(E-{3q^2\over 4m})\{Z^{(2)}_{c,c'}(q,q';E)+Z^{(3)}_{c,c'}(q,q';E)\}\tau_{c'}(E-{3q'^2\over 4m})g_{c'}(p')
\end{eqnarray}
with
\begin{eqnarray}
Z^{(2)}_{c,c'}(q,q';E)= \int _{-1}^1 { g_c(p(q,q'))g_{c'}(p'(q',q)) \over E-(q^2+q'^2+qq'x)/m+i\epsilon}P_{\cal L}(x) dx\Delta_{cc'}
\label{Z2}
\end{eqnarray}
and 
\begin{eqnarray}
&&Z^{(3)}_{c,c'}(q,q';E)=\int _0^\infty { g_c(p)h(p,q) \over E-(p^2+3q^2/4)/m+i\epsilon} {p^2dp \over 2\pi^2}
~~\tau_4(E)\cr 
&&\times \int _0^\infty { h(p',q') g_{c'}(p') \over E-(p'^2+3q'^2/4)/m+i\epsilon} {p'^2dp' \over 2\pi^2}\delta_{cc'}
\delta _{{\cal L}0} (1+2\Delta_{cc'})
\label{Z3}
\end{eqnarray}
where $P_{\cal L}(x)$ is the  Legendre polynomial.
Here $\Delta _{cc'}$ is the spin and isospin recoupling coefficient 
defined in the Appendix. We would like to emphasize that 
the quantities $Z^{(2)}$ and $Z^{(3)}$ can be calculated analytically (see the Appendix for details).
If the $Z_{cc'}^{(3)}$ term is omitted,  Eq.(\ref{AL}) is brought back to the original AL form.

\section{Cross Section and Scattering Length of Nd elastic scattering}

\begin{figure}[!h]
\begin{minipage}{0.45\hsize}
\begin{center}
\includegraphics[width=0.7\columnwidth,angle=0]{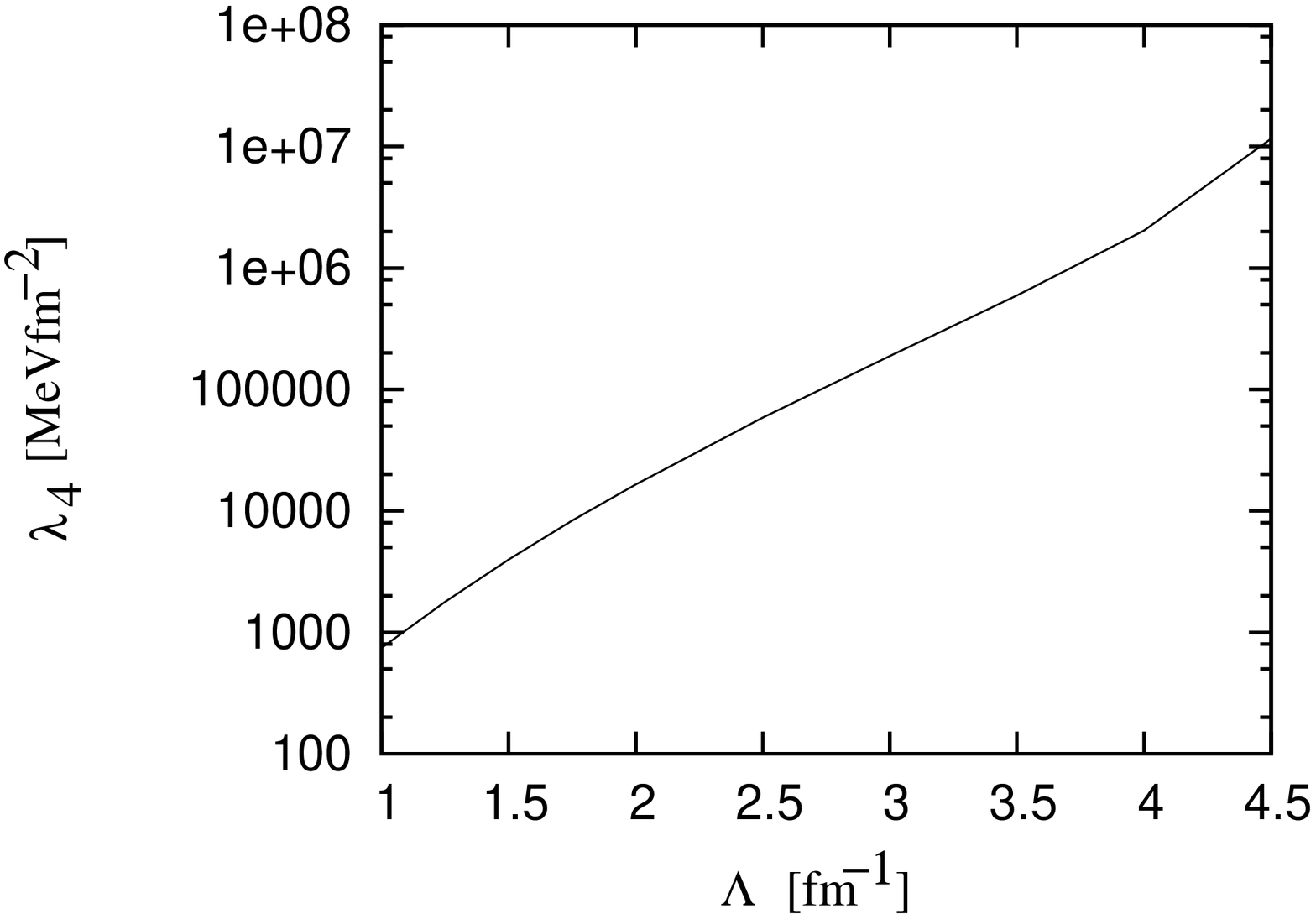} 
\caption{The relation between the parameters $\Lambda$ 
and $\lambda_4$ from Eq. (\ref{V4}), which is obtained by fixing the three-nucleon 
binding energy to $E_b$= -8.48 MeV. }
\label{fig1} 
\end{center}
\end{minipage}
\begin{minipage}{0.05\hsize}
\hspace{0.3cm}
\end{minipage}
\begin{minipage}{0.45\hsize}
\begin{center}
\includegraphics[width=0.7\columnwidth,angle=0]{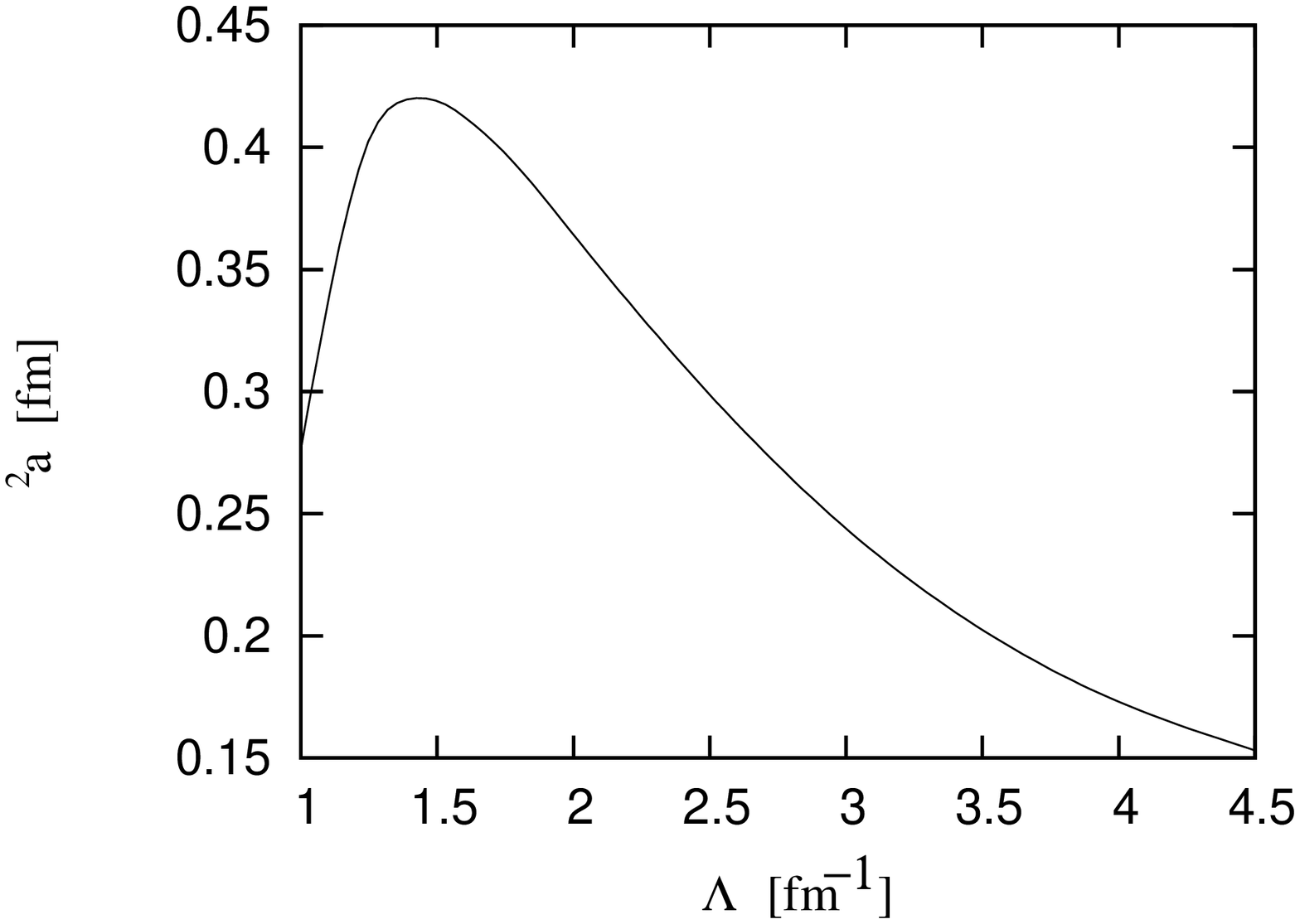}
\end{center}
\caption{The doublet scattering length $^2a$ as a function of the parameter $\Lambda$. Note 
that $\lambda_4$ and $\Lambda$ are now correlated as shown in Fig. \ref{fig1}.
The experimental value of $^2a$  is 0.65 $\pm$0.04 fm \cite{scatteringlength}.}
\label{fig2} 
\end{minipage}
\end{figure}

Here we would like to demonstrate some results.
The triton binding energy  calculated with the Yamaguchi potential \cite{Yamaguchi} is  -13.0 MeV.
Comparing this number with the experimental result (-8.48 MeV) clearly 
shows overbinding. 
This is because the Yamaguchi potential is too simple and furthermore 
restricted  
 to act only in S-waves.  Corresponding numbers obtained with modern potentials
 show in turn underbinding by about 1 MeV.
Since  our aim is to check consistency between  the numerical results obtained 
with  Eq. (\ref{Eq1}) and 
Eqs.(\ref{BKEq1})-(\ref{Eq4}), we introduce a simple 3NF as given in 
Eq. (\ref{V4}). 
 That 3NF contains two parameters ($\lambda_4$ and $\Lambda$), 
which are chosen to reproduce the experimental triton 
binding energy 8.48 MeV. Figure \ref{fig1} shows the relation between $\lambda_4$ and $\Lambda$ under this condition.  
Before introducing 3NF our result for the doublet scattering length is 
$^2a$ = -2.17 fm, which is very different  
from the experimental data,  0.65$\pm$0.04 fm \cite{scatteringlength}.
In the case of the quartet scattering length  our prediction is 
$^4a$ = 6.86 fm, which is rather close to the data (6.35$\pm$0.02fm)\cite{scatteringlength}.

Figure \ref{fig2} shows  results for $^2a$ when the 3NF is included. 
Though the doublet scattering length is sensitive to the 
strength of the 3NF, within the chosen range of the $\Lambda$ parameter the 
experimental value can not be reached. 
However, inclusion of the 3NF clearly improves the situation. 
On the other hand, the quartet scattering length $^4a$ hardly changes with 
$\Lambda$ and therefore is not displayed.

In the actual calculation we take $\Lambda$ = 2 fm$^{-1}$ and $\lambda_4$=16520  MeVfm$^{-2}$.
The Nd elastic scattering differential cross section is calculated at $E_{lab}$=14.1 MeV,
using  Eq. (\ref{Eq1}) and then separately   
with Eqs.(\ref{BKEq1})-(\ref{Eq4}). The Coulomb force is neglected in both our 
calculations. In order to achieve convergence,  all 3N states with total angular momentum
up to $31/2$ for the both parities have been included.
We use 100 $q$ Gaussian integral points and 20 $x$ points for the 
angular integration.
 The maximum value  of the $q$ momentum is set to  $q_{max}$= 20 fm$^{-1}$.
The smallest $\epsilon$, which appears in the Green's functions
in the complex energy method\cite{Kamada} is typically 0.01 fm$^{-1}$.
The approximation $t_4 \approx V_4$ provides elastic nucleon-deuteron cross
sections at forward angles which are by about 3-4 \% smaller than obtained
with $t_4$.

\begin{figure}[!h]
\centering
\includegraphics[width=0.35 \columnwidth,angle=0]{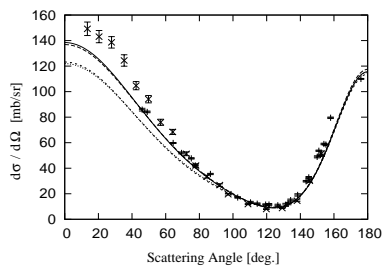}
\caption{The differential cross section of elastic Nd scattering at $E_{lab}$=14.1MeV. 
The short-dashed (dotted) line shows the theoretical results of the separable scheme (BK scheme) without 3NF.  
The solid (long-dashed) line demonstrates the results of the separable scheme (BK scheme) with 
the inclusion of 3NF. The cross "+" ("x") marks 
are for the experimental data from \cite{exp1} (\cite{exp2}). }
\label{fig3} 
\end{figure}

In Fig. \ref{fig3} the differential cross section for elastic Nd scattering 
is demonstrated,  comparing calculations with and without the 3NF. 
The theoretical predictions obtained with the two different schemes agree 
very well with each other (up to 1 \%) but, due to a use of very simple forces, 
yield only a fair 
description of the experimental data\cite{exp1,exp2}.

\section{Conclusions}

The aim of the present investigation was twofold. 
First of all we wanted to compare two calculational schemes, which deal with 
Nd scattering under the inclusion of a 3NF. 
To this aim we chose a simple separable form of the nucleon-nucleon potential 
and the 3NF. In this case, in the scheme using Eq. (\ref{Eq1}), 
some parts of the calculations can be done
analytically, which leads to very accurate results. 
The comparison of these results with the predictions based 
on Eqs. (\ref{BKEq1})--(\ref{Eq4}) provides a very good test for the numerical
performance of this second method, which treats directly any nucleon-nucleon 
and 3N forces. 
Secondly, since the first numerical framework uses the complex energy method,
our results provide a further example 
that this method can be successfully used in the few-nucleon
calculations. This is very encouraging in view of our planned 
applications of the complex energy method in the three-dimensional treatment
of two- and three-nucleon systems\cite{Golak}.

{\bf Acknowledgements:} This work was supported by the Polish National Science Center 
 under Grant No. DEC-2011/01/B/ST2/00578. 
 The numerical calculations were performed on the interactive server at 
RCNP, Osaka University, and on the supercomputer cluster of the JSC, J\"ulich, Germany.

\appendix \section{Analytical expressions}
 We show some analytical expressions used in the present investigation.
The 3NF of Eq.(\ref{V4})  satisfies 
\begin{eqnarray}
&&t_4(p,q,p',q';E)=V_4(p,q,p',q')\cr
&&+\int_0^\infty {p''^2dp \over 2\pi^2} \int _0 ^\infty {q''^2dq \over 2\pi^2} 
{V_4(p,q,p'',q'')t_4(p'',q'',p',q';E) \over E -(p''^2+3q''^2/4)/m + i \epsilon} .
\end{eqnarray}
Assuming the separable form ($t_4(p,q,p',q';E)=h(p,q)\tau_4(E)h(p',q')$) we have
$\tau_4(E)=\lambda_4+\lambda_4 \tau_4(E) I_4(E) \,$
where the integration  $I_4$ is performed using the so-called "hyper coordinates" ($\rho^2 =p^2+3q^2/4, \tan \theta=\sqrt{3}q/(2p)$),
\begin{eqnarray}
&&I_4(E)\equiv \int _0^\infty \int _0 ^\infty {h(p,q)^2\over E-(p^2+3q^2/4)/m + i \epsilon} {p^2 dp \over 2\pi^2}
{q^2 dq \over 2 \pi^2} \cr
&&={2 \over \sqrt{3} } {1 \over 4\pi^4} 
\int_ 0^\infty {\rho^5 d\rho \over (E-\rho^2/m+i\epsilon)
(\rho^2+\Lambda^2)^4}  \int _0^{\pi/2} \sin^2\theta \cos^2\theta d\theta \cr
&&
={m\over 384 \sqrt{3}\pi^3 \Lambda^2(Em+\Lambda^2)^4    } \cr
&& \times \left( 
{(Em +\Lambda^2)(2E^2m^2-5Em\Lambda^2-\Lambda^4)+6\Lambda^2E^2m^2
(\log(-Em)-\log(\Lambda^2))}\right)~~
\end{eqnarray}
where $\log(-Em)=\log(Em)-\pi i$ in case of $E>0$.
 In order to calculate the $Z_{cc'}^{(3)}$ terms we use
\begin{eqnarray}
&& \int _0^\infty {p^2dp\over 2\pi^2} {h(p,q)g(p)\over E-(p^2+3q^2/4)/m + i \epsilon} 
=-{m(\beta-ik'+2\Lambda')\over 8\pi
(\beta-i k')\Lambda'(\beta+\Lambda')^2(\Lambda'-ik')^2},~~
\end{eqnarray}
with $k'\equiv \sqrt{mE-3q^2/4}$ and $\Lambda'\equiv\sqrt{3q^2/4+\Lambda^2}$.
 
Finally, 
the recoupling coefficients $\Delta_{cc'}$ for spin doublet ($I=\frac12 $) and quartet 
($I=\frac32$) states in Eqs.(\ref{Z2})-(\ref{Z3}) are 
\begin{eqnarray}
\Delta_{cc'}^{I=\frac12}=\left(
\begin{array}{cc}
\Delta_{dd} & \Delta_{d\phi} \cr
\Delta_{\phi d} & \Delta_{\phi\phi} \cr 
\end{array}
 \right)_{I=1/2}=
\left(
\begin{array}{cc}
\frac14& -\frac34 \cr
-\frac34 & \frac14 \cr 
\end{array}
\right),
~~~\Delta_{cc'}^{I=\frac32}=
\left(
\begin{array}{cc}
-\frac12& 0 \cr
0 & 0 \cr 
\end{array}
\right).~~~
\end{eqnarray}

\end{document}